\documentclass[journal]{IEEEtran}
\usepackage{amsmath,amsfonts}
\usepackage{amsthm}

\usepackage{amssymb}
\usepackage{mathtools}  
\usepackage{algorithmic}
\usepackage{array}
\usepackage[caption=false,font=normalsize,labelfont=sf,textfont=sf]{subfig}
\usepackage{textcomp}
\usepackage{stfloats}
\usepackage{cite}
\usepackage{url}
\usepackage{verbatim}
\usepackage{graphicx}
\def\BibTeX{{\rm B\kern-.05em{\sc i\kern-.025em b}\kern-.08em
  T\kern-.1667em\lower.7ex\hbox{E}\kern-.125emX}}
\usepackage{balance}
\begin{document}
\title{Competent Discrete Time Modeling For analogue controlled PWM Converter Considering State-Feedback}
\author{Yuxin Yang, Hang Zhou, Hourong Song, Branislav Hredzak
\thanks{This paper is intended to be submitted to IEEE Transactions on Power Electronics}}

\markboth{Letter IEEE Transactions on Power Electronics}%
{Discrete Time modeling for analogue controlled DC-DC converters}

\maketitle

\begin{abstract}
Ever since R.D.Middlebrook proposed the state space averaging notion. The small signal model has been widely used as a design tool to tune control parameters. As Moore's law is continuing and the AI chip's high demand for power consumption and dynamic response, the control bandwidth needs to be boosted. However, the average model has two basic assumptions: the low-frequency assumption, the small ripple assumption. In high-bandwidth design, these two assumptions are violated. In order to solve this, various methods have been proposed. This paper gives a comprehensive overview of the existing small signal model for PWM converters from the following perspectives: 1. model fidelity, 2. analytical tractability. 3. complexity of the derivation process and result 4.generality.
\end{abstract}


\section{Introduction}
\IEEEPARstart{P}{WM} converters are the fundamental part of power conversion since the American scholar William.E.Newell proposed the notion "Power Electronics". A Small Signal Linearized Model is required to design a quick and robust dynamic system. The state space averaging method is first proposed\cite{4421} \cite{CUK_SSA}. The slide-averaging operator is introduced to simplify the model. However, two important assumptions are introduced:
\begin{itemize}
    \item Small-Ripple Assumption
    The switching ripple is well-attenuated in the PWM control system input. Thus, the PWM can be regarded as a pure gain term.
    \item Low-frequency Assumption
    The control bandwidth is a lot lower than the switching frequency. Therefore, the side-band components in the system is neglected.
\end{itemize}
In the late 1970s and 1990s, the ripple-based control methods\cite{PCM_notion} \cite{Goder1996V2} were introduced to enhance the dynamic response. Moreover, the high bandwidth Average-Mode Controlled converters are introduced. Therefore, two effects become significant:
\begin{itemize}
\item Sideband (sampling) Effect: As shown on the figure, the sideband effect exists in the system, since the PWM modulator's small signal behavior is equivalent to a sampler (dirac comb). As the bandwidth is boosted, the high-frequency components are injected. Then, the low-frequency assumption imposes significant restrictions.
\item Ripple Effect
As less attenuation is introduced into the control system, the waveform in the comparator input has more side-band components than in the case of low-bandwidth average mode control. Then the low-ripple assumption imposes significant restrictions.
\end{itemize}
In the ripple-based controlled PWM converters, the two assumptions in the SSA methods are violated. In order to model the ripple-based control, various models are proposed. Either or both of these two effects are considered.
In the left part of the Introduction, a brief review on the modelling method will be shown.
\subsection{Multi-frequency Method}
\subsubsection{Two-frequency Model}
\label{susubsection_Two-freq}
The Fourier-Integral based describing function method is used in \cite{yangqiu_multi} to describe the frequency-coupling behavior.  Therefore, the sideband-effect is considered. However, since only the lower-sideband ($fsw+fp$) generated by switching frequency and perturbation frequency is considered, it cannot explain the side-band effect competently.  The ripple effect in the system is ignored. In terms of the analytical tractability, since the sidebands are introduced in the form of an iteration denominator, the model cannot show the analytical symbolic stability boundary. The derivation process is very complex.
\subsubsection{Four-fequency Model}
\label{susubsection_Four-freq}
This model is very similar to the Two-frequency Model. More sidebands are considered. And the ripple effect in modulation is considered. Because of more considered sidebands, it is more accurate. However, it is still not competent to fully explain the sideband effect. Moreover, as more sidebands are considered, the complexity of the derivation process of these types of Fourier-Integral-based methods increases significantly. This phenomena also happens on its result, which also further diminishes its analytical tractability compared to the Two-frequency Model. 
\subsubsection{Matrix-Based Multi-Frequency Model (HSS)}
The derivation process of the Fourier integral in \cite{yangqiu_multi} and \cite{hsiao_fourfreq} is very complex. However, the sideband effect requires more sidebands to be considered to enhance the model fidelity. Based on the sampling theorem (the Poisson summation formulae), the sideband mapping relation is derived easily without a complex Fourier integral. Furthermore, the derivation process is also simplified using the linear algebra tool. Therefore, the derivation process and the result of the model are expressed in a simpler form. Introducing arbitary numbers of sidebands become possible under this improvement. Moreover, the modeling result is extended to the MIMO form. The interaction (such as the beat frequency oscillation) between converters that are cascade connected can be explained by the model.\\
However, this model's improvement only simplified the derivation process of the model. The model result is still in a complex iteration form. The analytical tractability of the model results is diminished even more.
\subsection{Sampled-data Method}
\subsection{Closed-loop method}
\subsubsection{Closed-loop discrete model with approximation}
In \cite{ridley_continuous}, the digital small signal perturbation analysis is applied to the PCM-controlled BUCK converter. The voltage-second balancing approximation is introduced.
In order to make the model more understandable, the pade approximation is introduced. Consequently, the model form is a simple Laplace-domain polynomial. So, the model form is very friendly to industry utilization. It is the first model that can help to predict sub-harmonic oscillations in a PCM-controlled converter effectively.  The simplification using the Pade approximation enables the model to transfer from exponential(the Z operator) and polynomial hybrid form to a Laplace domain pure polynomial form. This class of models has gained wide acceptance among engineers, owing to its simple structure and Laplace-domain formulation, which align with conventional engineering practice. The derivation process of this class of models is often less rigorous than desired, and at times seems insufficiently motivated, especially for the introduction of ZOH. This limitation poses challenges for systematic extension and raises concerns regarding the strength of the underlying theoretical framework.
\subsubsection{Closed-loop Describing Function}
This model \cite{jianli_model_approach} utilizes the same mathematical tools as the model in section \ref {susubsection_Two-freq}. It focus on the time domain closed-loop relationship, therefore both the ripple and sideband effect are considered. In the derivation process, the author introduced the voltage-second balance approximation in \cite{jianli_model_approach} and ampere-second approximation in $V^2$ extension \cite{jianli_V2}. This category of models inherits the Pade approximation framework, thus preserving the the same practical advantages of simplicity and familiarity to engineers as the Closed-loop discrete model. In terms of the the forms of the result, the system is expressed as a polynomial in the Laplace domain. This representation also incorporates the concept of the linear two-port equivalent circuit, which was widely referenced during the era of averaged models, as a means of presenting the model results. Although this equivalent circuit provides a certain degree of physical intuition, it is in fact introduced by fitting the model output rather than being directly observed from the modeling process. The underlying derivation remains essentially a perturbation-based approach under closed-loop conditions. Unlike the Closed-loop discrete model, the derivation here is conducted with much greater rigor. This enhanced rigor makes it possible to extend the modeling framework without exposing the subsequent developments to the theoretical vulnerabilities of the Closed-loop discrete model. However, the derivation process of this type of model is still highly complex.
\section{Analytical Steady State modelling for converters using Poincaré map}
We consider a piecewise linear time‐invariant system over one period \(T_s\), divided into \(n\) consecutive intervals of durations \(T_1,\dots,T_n\) (with \(T_s=\sum_{i=1}^n T_i\)).  On interval \(i\) the system is governed by
\[
\dot{\mathbf X}(t)
= A_i\,\mathbf X(t) + B_i\,\mathbf U,
\]
where the input \(\mathbf U\) is held constant in each interval.  Denote the state at the start of the \(i\)th interval by \(\mathbf X_{i-1}\), then the discrete‐time update is
\begin{equation}
    \begin{aligned}
        &\mathbf X_i
        = \Phi_i\,\mathbf X_{i-1} \;+\;\Gamma_i,
        \quad
        \Phi_i = e^{A_iT_i},\\
        &\Gamma_i = \int_0^{T_i} e^{A_i(T_i-\tau)}\,B_i\,\mathbf U\,d\tau.
    \end{aligned}
\end{equation}
If $A_i$ is invertible, 
\begin{equation}
\begin{aligned}
    &\Gamma_i = \int_0^{T_i} e^{A_i(T_i-\tau)}\,B_i\,\mathbf U\,d\tau\\&=
    \int_0^T e^{A\tau}B_iU\,d\tau
    = A_i^{-1}\bigl(e^{A_iT}-I\bigr)B_iU.
\end{aligned}
\end{equation}
Our goal is to derive:
\begin{enumerate}
  \item A closed‐form expression for \(\mathbf X_n\) in terms of \(\mathbf X_0\) and the \(\Gamma_i\).
  \item The fixed‐point equation for the periodic steady state \(\mathbf X^*\) satisfying \(\mathbf X_n=\mathbf X_0=\mathbf X^*\).
\end{enumerate}
\section{Product Notation: Direction and Boundary}

We introduce two notations for multiplying the transition matrices:
\[
\prod_{j=a}^{b}\!{}^{\rightarrow}\Phi_j
\;:=\;\Phi_a\,\Phi_{a+1}\,\cdots\,\Phi_b,
\quad
\prod_{j=a}^{b}\!{}^{\leftarrow}\Phi_j
\;:=\;\Phi_b\,\Phi_{b-1}\,\cdots\,\Phi_a,
\]
for \(a\le b\).  In particular,
\[
\prod_{j=a}^{a}\!{}^{\rightarrow}\Phi_j
= \prod_{j=a}^{a}\!{}^{\leftarrow}\Phi_j
= \Phi_a.
\]
In our closed‐form formula only the “reverse” product \(\prod_{j=1}^{n}{}^{\leftarrow}\Phi_j\) appears, avoiding any need for an empty‐product convention.

\section{Main Results: Recursive Closed‐Form and Fixed‐Point Equation}

\subsection{Recursive Closed‐Form}

For any \(n\ge1\), the state at the end of the \(n\)th interval is
\[
\boxed{
\mathbf X_n
= \Bigl(\prod_{j=1}^{n}{}^{\leftarrow}\Phi_j\Bigr)\,\mathbf X_0
+ \sum_{i=1}^{n-1}
  \Bigl(\prod_{j=i+1}^{n}{}^{\leftarrow}\Phi_j\Bigr)\,\Gamma_i
+ \Gamma_n
}
\tag{1}
\]

\subsection{Periodic Fixed‐Point Equation}

If a periodic steady state \(\mathbf X^*=\mathbf X_0=\mathbf X_n\) exists, it satisfies
\[
\boxed{
\bigl(I - \Pi\bigr)\,\mathbf X^*
= \sum_{i=1}^{n-1}
  \Bigl(\prod_{j=i+1}^{\,n}{}^{\leftarrow}\Phi_j\Bigr)\,\Gamma_i
+ \Gamma_n,
}
\quad
\Pi := \prod_{j=1}^{n}{}^{\leftarrow}\Phi_j.
\tag{2}
\]
Take four state transition as an example:
\begin{equation}
\begin{aligned}
X_1 &= \Phi_1 X_0 + \Gamma_1,\\
X_2 &= \Phi_2\Phi_1 X_0 + \Phi_2\,\Gamma_1 + \Gamma_2,\\
X_3 &= \Phi_3\Phi_2\Phi_1 X_0 + \Phi_3\Phi_2\,\Gamma_1 + \Phi_3\,\Gamma_2 + \Gamma_3,\\
X_4 &= \Phi_4\Phi_3\Phi_2\Phi_1 X_0 
      + \Phi_4\Phi_3\Phi_2\,\Gamma_1 
      + \Phi_4\Phi_3\,\Gamma_2 
      + \Phi_4\,\Gamma_3 
      + \Gamma_4.
\end{aligned}
\end{equation}

\section{Jacobian Linearization of PWL Switching Maps Under Four PWM Logics}
\label{sec:pwl_jacobian_four_pwm}

This section formulates a unified Jacobian-based linearization for piecewise-linear (PWL) switching systems under four PWM logics: (i) constant-on-time (COT), (ii) constant-off-time (COFT), (iii) fixed-frequency trailing-edge (peak) PWM, and (iv) fixed-frequency leading-edge (valley) PWM. The differences among these cases are (a) the sampling phase (peak vs.\ valley), (b) the subinterval ordering in the Poincar\'e map, and (c) which subinterval durations are fixed or time-varying.

\subsubsection*{A. PWL explicit solution and timing derivatives}
Consider a PWL segment $i$ governed by
\begin{equation}
\dot{x}(t)=A_i x(t)+B_iU,
\label{eq:pwl_segment}
\end{equation}
where $x\in\mathbb{R}^n$ and $U$ is constant over the segment. The segment endpoint mapping over duration $T$ is
\begin{equation}
\begin{aligned}
x(T) &= \Phi_i(T)\,x(0)+\Gamma_i(T),\\
\Phi_i(T) &:= e^{A_iT},\\
\Gamma_i(T) &:= \int_{0}^{T} e^{A_i(T-\tau)}B_iU\,d\tau .
\end{aligned}
\label{eq:pwl_explicit}
\end{equation}
If $A_i$ is invertible, $\Gamma_i(T)=A_i^{-1}\big(\Phi_i(T)-I\big)B_iU$ may be used; otherwise the integral definition in \eqref{eq:pwl_explicit} remains valid.

Time derivatives required for Jacobian linearization are
\begin{equation}
\begin{aligned}
\frac{\partial \Phi_i(T)}{\partial T} &= A_i\Phi_i(T)=\Phi_i(T)A_i,\\
\frac{\partial \Gamma_i(T)}{\partial T} &= \Phi_i(T)B_iU .
\end{aligned}
\label{eq:PhiGamma_time_derivatives}
\end{equation}
These identities eliminate manual perturbation expansions of matrix exponentials.

\subsubsection*{B. Unified event (comparator) timing relation}
Let the switching event be determined by a comparator quantity $y(t)=Kx(t)$ intersecting a ramp of slope $S_e$ against a control voltage $v_c$. At the switching edge,
\begin{equation}
Kx_{\mathrm{edge}}=v_c + S_e\,\Delta T,
\label{eq:event_equation}
\end{equation}
where $\Delta T$ denotes the relevant timing variable (e.g., $T_{\mathrm{on}}$ or $T_{\mathrm{off}}$). Linearizing \eqref{eq:event_equation} yields
\begin{equation}
\widehat{\Delta T}
=\frac{K\hat{x}_{\mathrm{edge}}-\hat{v}_c}{S_e}.
\label{eq:event_linear}
\end{equation}
In the four cases below, the sampling instant is chosen at the switching edge so that $x_{\mathrm{edge}}$ coincides with the sampled state, enabling elimination of $\widehat{\Delta T}$ inside the linearized map.

\subsection{Constant-On-Time (COT): Valley-Triggered, Fixed $T_{\mathrm{on}}$}
\label{subsec:COT}

In COT, the on-time $T_{\mathrm{on}}$ is fixed while the off-time $T_{\mathrm{off},k}$ varies and is determined by the next valley-trigger event. Sampling is taken at the valley (turn-on) instant. Over one cycle,
\[
\text{on (fixed }T_{\mathrm{on}})\ \rightarrow\ \text{off (variable }T_{\mathrm{off},k}).
\]

\subsubsection*{A. Large-signal Poincar\'e map}
Let $x_k$ denote the sampled state at the valley of cycle $k$. The one-step map is
\begin{equation}
\begin{aligned}
x_{k+1}
&=\Phi_{\mathrm{off}}(T_{\mathrm{off},k})
\Big(
\Phi_{\mathrm{on}}(T_{\mathrm{on}})x_k
+\Gamma_{\mathrm{on}}(T_{\mathrm{on}})
\Big)
+\Gamma_{\mathrm{off}}(T_{\mathrm{off},k}).
\end{aligned}
\label{eq:cot_map}
\end{equation}

\subsubsection*{B. Jacobian blocks}
At the periodic steady state $(x^\star,T_{\mathrm{off}}^\star)$, define
\begin{equation}
\begin{aligned}
\Phi_{\mathrm{on}} &:= \Phi_{\mathrm{on}}(T_{\mathrm{on}}),\\
\Phi_{\mathrm{off}}^\star &:= \Phi_{\mathrm{off}}(T_{\mathrm{off}}^\star),\\
\Gamma_{\mathrm{on}} &:= \Gamma_{\mathrm{on}}(T_{\mathrm{on}}),\\
\Gamma_{\mathrm{off}}^\star &:= \Gamma_{\mathrm{off}}(T_{\mathrm{off}}^\star).
\end{aligned}
\label{eq:cot_defs}
\end{equation}
The state Jacobian is
\begin{equation}
\phi_{\mathrm{COT}}
=\left.\frac{\partial x_{k+1}}{\partial x_k}\right|_{\star}
=\Phi_{\mathrm{off}}^\star\,\Phi_{\mathrm{on}}.
\label{eq:cot_phi}
\end{equation}
Using \eqref{eq:PhiGamma_time_derivatives}, the off-time sensitivity is
\begin{equation}
\begin{aligned}
\gamma_{\mathrm{off}}
&=\left.\frac{\partial x_{k+1}}{\partial T_{\mathrm{off}}}\right|_{\star}\\
&=A_{\mathrm{off}}\Phi_{\mathrm{off}}^\star
\Big(\Phi_{\mathrm{on}}x^\star+\Gamma_{\mathrm{on}}\Big)
+\Phi_{\mathrm{off}}^\star B_{\mathrm{off}}U .
\end{aligned}
\label{eq:cot_gamma_off}
\end{equation}

\subsubsection*{C. Eliminating $\widehat{T}_{\mathrm{off},k}$}
The valley event occurs at the sampling instant, hence $x_{\mathrm{edge}}=x_{k+1}$ and $\Delta T=T_{\mathrm{off},k}$. Linearizing \eqref{eq:cot_map} gives
\begin{equation}
\hat{x}_{k+1}
=\phi_{\mathrm{COT}}\hat{x}_k
+\Gamma_{\mathrm{off}}\widehat{T}_{\mathrm{off},k}.
\label{eq:cot_lin_raw}
\end{equation}
Substituting \eqref{eq:event_linear} yields
\begin{equation}
\begin{aligned}
\Big(I-\Gamma_{\mathrm{off}}K/S_e\Big)\hat{x}_{k+1}
&=\Phi_{\mathrm{COT}}\hat{x}_k
-\Gamma_{\mathrm{off}}\hat{v}_{c,k+1}/S_e .
\end{aligned}
\label{eq:cot_lin_final}
\end{equation}

\subsection{Constant-Off-Time (COFT): Peak-Triggered, Fixed $T_{\mathrm{off}}$}
\label{subsec:COFT}

In COFT, the off-time $T_{\mathrm{off}}$ is fixed while the on-time $T_{\mathrm{on},k}$ varies and is determined by the next peak-trigger event. Sampling is taken at the peak (turn-off) instant. Over one cycle,
\[
\text{off (fixed }T_{\mathrm{off}})\ \rightarrow\ \text{on (variable }T_{\mathrm{on},k}).
\]

\subsubsection*{A. Large-signal Poincar\'e map}
Let $x_k$ denote the sampled state at the peak of cycle $k$. The one-step map is
\begin{equation}
\begin{aligned}
x_{k+1}
&=\Phi_{\mathrm{on}}(T_{\mathrm{on},k})
\Big(
\Phi_{\mathrm{off}}(T_{\mathrm{off}})x_k
+\Gamma_{\mathrm{off}}(T_{\mathrm{off}})
\Big)
+\Gamma_{\mathrm{on}}(T_{\mathrm{on},k}).
\end{aligned}
\label{eq:coft_map}
\end{equation}

\subsubsection*{B. Jacobian blocks}
At the periodic steady state $(x^\star,T_{\mathrm{on}}^\star)$, define
\begin{equation}
\begin{aligned}
\Phi_{\mathrm{off}} &:= \Phi_{\mathrm{off}}(T_{\mathrm{off}}),\\
\Phi_{\mathrm{on}}^\star &:= \Phi_{\mathrm{on}}(T_{\mathrm{on}}^\star),\\
\Gamma_{\mathrm{off}} &:= \Gamma_{\mathrm{off}}(T_{\mathrm{off}}),\\
\Gamma_{\mathrm{on}}^\star &:= \Gamma_{\mathrm{on}}(T_{\mathrm{on}}^\star).
\end{aligned}
\label{eq:coft_defs}
\end{equation}
The state Jacobian is
\begin{equation}
\phi_{\mathrm{COFT}}
=\left.\frac{\partial x_{k+1}}{\partial x_k}\right|_{\star}
=\Phi_{\mathrm{on}}^\star\,\Phi_{\mathrm{off}}.
\label{eq:coft_phi}
\end{equation}
The on-time sensitivity is
\begin{equation}
\begin{aligned}
\gamma_{\mathrm{on}}
&=\left.\frac{\partial x_{k+1}}{\partial T_{\mathrm{on}}}\right|_{\star}\\
&=A_{\mathrm{on}}\Phi_{\mathrm{on}}^\star
\Big(\Phi_{\mathrm{off}}x^\star+\Gamma_{\mathrm{off}}\Big)
+\Phi_{\mathrm{on}}^\star B_{\mathrm{on}}U .
\end{aligned}
\label{eq:coft_gamma_on}
\end{equation}

\subsubsection*{C. Eliminating $\widehat{T}_{\mathrm{on},k}$}
The peak event occurs at the sampling instant, hence $x_{\mathrm{edge}}=x_{k+1}$ and $\Delta T=T_{\mathrm{on},k}$. Linearizing \eqref{eq:coft_map} gives
\begin{equation}
\hat{x}_{k+1}
=\phi_{\mathrm{COFT}}\hat{x}_k
+\Gamma_{\mathrm{on}}\widehat{T}_{\mathrm{on},k}.
\label{eq:coft_lin_raw}
\end{equation}
Substituting \eqref{eq:event_linear} yields
\begin{equation}
\begin{aligned}
\Big(I-\Gamma_{\mathrm{on}}K/S_e\Big)\hat{x}_{k+1}
&=\phi_{\mathrm{COFT}}\hat{x}_k
-\Gamma_{\mathrm{on}}\hat{v}_{c,k+1}/S_e .
\end{aligned}
\label{eq:coft_lin_final}
\end{equation}

\subsection{Fixed-Frequency Trailing-Edge (Peak) PWM}
\label{subsec:FF_peak}

In fixed-frequency trailing-edge PWM, the switch is turned on by the clock
at the beginning of each cycle, while the comparator triggers turn-off
within the cycle. The switching period is constant, hence for every cycle $k$,
\begin{equation}
T_{\mathrm{on},k}+T_{\mathrm{off},k}=T_s,
\qquad
T_{\mathrm{off},k}=T_s-T_{\mathrm{on},k}.
\label{eq:ff_trailing_constraint}
\end{equation}
Let the ramp have amplitude $V_r$ over one period $T_s$, so its slope is
\begin{equation}
S_e:=\frac{V_r}{T_s}.
\label{eq:ff_ramp_slope}
\end{equation}

\subsubsection{Large-signal peak-to-peak map}
\label{subsubsec:ff_peak_map}

Let $x_k$ be sampled at the peak (turn-off) edge of cycle $k$.
From this sampling instant to the next peak, the trajectory consists of
\emph{(i)} the off-interval of cycle $k$ with duration $T_{\mathrm{off},k}$,
followed by \emph{(ii)} the on-interval of cycle $k{+}1$ with duration
$T_{\mathrm{on},k+1}$. Therefore,
\begin{equation}
\begin{aligned}
x_{k+1}
&=\Phi_{\mathrm{on}}\!\big(T_{\mathrm{on},k+1}\big)
\Big(
\Phi_{\mathrm{off}}\!\big(T_{\mathrm{off},k}\big)x_k
+\Gamma_{\mathrm{off}}\!\big(T_{\mathrm{off},k}\big)
\Big)\\
&\quad+\Gamma_{\mathrm{on}}\!\big(T_{\mathrm{on},k+1}\big),
\end{aligned}
\label{eq:ff_peak_map}
\end{equation}
where, for each mode $m\in\{\mathrm{on},\mathrm{off}\}$,
\(
\Phi_m(T)=e^{A_m T}
\)
and
\(
\Gamma_m(T)=\int_0^T e^{A_m(T-\tau)}B_mU\,d\tau
\).

\subsubsection{Jacobian blocks with two timing perturbations}
\label{subsubsec:ff_peak_jacobian}

At the steady state, denote
\begin{equation}
T_{\mathrm{on}}^\star,\quad
T_{\mathrm{off}}^\star:=T_s-T_{\mathrm{on}}^\star,
\qquad
x^\star.
\end{equation}
Define
\begin{equation}
\Phi_{\mathrm{on}}^\star:=\Phi_{\mathrm{on}}(T_{\mathrm{on}}^\star),\quad
\Phi_{\mathrm{off}}^\star:=\Phi_{\mathrm{off}}(T_{\mathrm{off}}^\star),\quad
\Gamma_{\mathrm{off}}^\star:=\Gamma_{\mathrm{off}}(T_{\mathrm{off}}^\star).
\label{eq:ff_peak_defs}
\end{equation}
The state Jacobian is
\begin{equation}
\Phi_{\mathrm{FF,pk}}
=\left.\frac{\partial x_{k+1}}{\partial x_k}\right|_{\star}
=\Phi_{\mathrm{on}}^\star\,\Phi_{\mathrm{off}}^\star .
\label{eq:ff_peak_phi}
\end{equation}

Define two timing sensitivities corresponding to the two durations that
appear in \eqref{eq:ff_peak_map}.
First, the sensitivity to the \emph{next-cycle} on-time $T_{\mathrm{on},k+1}$:
\begin{equation}
\begin{aligned}
\gamma_{+}
&:=\left.\frac{\partial x_{k+1}}{\partial T_{\mathrm{on},k+1}}\right|_{\star}\\
&=A_{\mathrm{on}}\Phi_{\mathrm{on}}^\star
\Big(\Phi_{\mathrm{off}}^\star x^\star+\Gamma_{\mathrm{off}}^\star\Big)
+\Phi_{\mathrm{on}}^\star B_{\mathrm{on}}U .
\end{aligned}
\label{eq:ff_peak_gamma_plus}
\end{equation}
Second, the sensitivity to the \emph{current-cycle} on-time $T_{\mathrm{on},k}$,
which enters \eqref{eq:ff_peak_map} only through
$T_{\mathrm{off},k}=T_s-T_{\mathrm{on},k}$ in \eqref{eq:ff_trailing_constraint}:
\begin{equation}
\begin{aligned}
\gamma_{-}
&:=\left.\frac{\partial x_{k+1}}{\partial T_{\mathrm{on},k}}\right|_{\star}
=-\left.\frac{\partial x_{k+1}}{\partial T_{\mathrm{off},k}}\right|_{\star}\\
&=-\Phi_{\mathrm{on}}^\star
\Big(
A_{\mathrm{off}}\Phi_{\mathrm{off}}^\star x^\star
+\Phi_{\mathrm{off}}^\star B_{\mathrm{off}}U
\Big).
\end{aligned}
\label{eq:ff_peak_gamma_minus}
\end{equation}
Thus the linearized map is
\begin{equation}
\hat{x}_{k+1}
=\Phi_{\mathrm{FF,pk}}\hat{x}_k
+\gamma_{+}\widehat{T}_{\mathrm{on},k+1}
+\gamma_{-}\widehat{T}_{\mathrm{on},k}.
\label{eq:ff_peak_lin_raw}
\end{equation}

\subsubsection{Eliminating $\widehat{T}_{\mathrm{on},k+1}$ at the peak edge}
\label{subsubsec:ff_peak_elim}

At the peak (turn-off) sampling instant, the edge time is
$\Delta T=T_{\mathrm{on},k+1}$ and $x_{\mathrm{edge}}=x_{k+1}$.
Using the standard ramp-comparator small-signal relation,
\begin{equation}
\widehat{T}_{\mathrm{on},k+1}
=\frac{K\hat{x}_{k+1}-\hat{v}_{c,k+1}}{S_e}.
\label{eq:ff_peak_event_sub}
\end{equation}
Substituting \eqref{eq:ff_peak_event_sub} into \eqref{eq:ff_peak_lin_raw}
yields
\begin{equation}
\begin{aligned}
\Big(I-\gamma_{+}K/S_e\Big)\hat{x}_{k+1}
&=\Phi_{\mathrm{FF,pk}}\hat{x}_k
+\gamma_{-}\widehat{T}_{\mathrm{on},k}
-\gamma_{+}\hat{v}_{c,k+1}/S_e .
\end{aligned}
\label{eq:ff_peak_lin_final}
\end{equation}

\subsection{Fixed-Frequency Leading-Edge (Valley) PWM}
\label{subsec:FF_valley}

In fixed-frequency leading-edge PWM, the clock forces turn-off at the
beginning of each cycle, while the comparator triggers turn-on (valley)
within the cycle. The fixed-period constraint implies, for every cycle $k$,
\begin{equation}
T_{\mathrm{on},k}+T_{\mathrm{off},k}=T_s,
\qquad
T_{\mathrm{on},k}=T_s-T_{\mathrm{off},k}.
\label{eq:ff_leading_constraint}
\end{equation}
The ramp slope $S_e$ is still given by \eqref{eq:ff_ramp_slope}.

\subsubsection{Large-signal valley-to-valley map}
\label{subsubsec:ff_valley_map}

Let $x_k$ be sampled at the valley (turn-on) edge of cycle $k$.
From this sampling instant to the next valley, the trajectory consists of
\emph{(i)} the on-interval of cycle $k$ with duration
$T_{\mathrm{on},k}=T_s-T_{\mathrm{off},k}$, followed by \emph{(ii)} the
off-interval of cycle $k{+}1$ with duration $T_{\mathrm{off},k+1}$.
Therefore,
\begin{equation}
\begin{aligned}
x_{k+1}
&=\Phi_{\mathrm{off}}\!\big(T_{\mathrm{off},k+1}\big)
\Big(
\Phi_{\mathrm{on}}\!\big(T_{\mathrm{on},k}\big)x_k
+\Gamma_{\mathrm{on}}\!\big(T_{\mathrm{on},k}\big)
\Big)\\
&\quad+\Gamma_{\mathrm{off}}\!\big(T_{\mathrm{off},k+1}\big),
\end{aligned}
\label{eq:ff_valley_map}
\end{equation}
with $T_{\mathrm{on},k}=T_s-T_{\mathrm{off},k}$ per
\eqref{eq:ff_leading_constraint}.

\subsubsection{Jacobian blocks with two timing perturbations}
\label{subsubsec:ff_valley_jacobian}

At the steady state, denote
\begin{equation}
T_{\mathrm{off}}^\star,\quad
T_{\mathrm{on}}^\star:=T_s-T_{\mathrm{off}}^\star,
\qquad
x^\star.
\end{equation}
Define
\begin{equation}
\Phi_{\mathrm{on}}^\star:=\Phi_{\mathrm{on}}(T_{\mathrm{on}}^\star),\quad
\Phi_{\mathrm{off}}^\star:=\Phi_{\mathrm{off}}(T_{\mathrm{off}}^\star),\quad
\Gamma_{\mathrm{on}}^\star:=\Gamma_{\mathrm{on}}(T_{\mathrm{on}}^\star).
\label{eq:ff_valley_defs}
\end{equation}
The state Jacobian is
\begin{equation}
\Phi_{\mathrm{FF,val}}
=\left.\frac{\partial x_{k+1}}{\partial x_k}\right|_{\star}
=\Phi_{\mathrm{off}}^\star\,\Phi_{\mathrm{on}}^\star .
\label{eq:ff_valley_phi}
\end{equation}

Define the two timing sensitivities.
First, the sensitivity to the \emph{next-cycle} off-time $T_{\mathrm{off},k+1}$:
\begin{equation}
\begin{aligned}
\gamma_{+}
&:=\left.\frac{\partial x_{k+1}}{\partial T_{\mathrm{off},k+1}}\right|_{\star}\\
&=A_{\mathrm{off}}\Phi_{\mathrm{off}}^\star
\Big(\Phi_{\mathrm{on}}^\star x^\star+\Gamma_{\mathrm{on}}^\star\Big)
+\Phi_{\mathrm{off}}^\star B_{\mathrm{off}}U .
\end{aligned}
\label{eq:ff_valley_gamma_plus}
\end{equation}
Second, the sensitivity to the \emph{current-cycle} off-time $T_{\mathrm{off},k}$
which enters \eqref{eq:ff_valley_map} only through
$T_{\mathrm{on},k}=T_s-T_{\mathrm{off},k}$ in \eqref{eq:ff_leading_constraint}:
\begin{equation}
\begin{aligned}
\gamma_{-}
&:=\left.\frac{\partial x_{k+1}}{\partial T_{\mathrm{off},k}}\right|_{\star}
=-\left.\frac{\partial x_{k+1}}{\partial T_{\mathrm{on},k}}\right|_{\star}\\
&=-\Phi_{\mathrm{off}}^\star
\Big(
A_{\mathrm{on}}\Phi_{\mathrm{on}}^\star x^\star
+\Phi_{\mathrm{on}}^\star B_{\mathrm{on}}U
\Big).
\end{aligned}
\label{eq:ff_valley_gamma_minus}
\end{equation}
Thus
\begin{equation}
\hat{x}_{k+1}
=\Phi_{\mathrm{FF,val}}\hat{x}_k
+\gamma_{+}\widehat{T}_{\mathrm{off},k+1}
+\gamma_{-}\widehat{T}_{\mathrm{off},k}.
\label{eq:ff_valley_lin_raw}
\end{equation}

\subsubsection{Eliminating $\widehat{T}_{\mathrm{off},k+1}$ at the valley edge}
\label{subsubsec:ff_valley_elim}

At the valley (turn-on) sampling instant, the edge time is
$\Delta T=T_{\mathrm{off},k+1}$ and $x_{\mathrm{edge}}=x_{k+1}$.
Thus
\begin{equation}
\widehat{T}_{\mathrm{off},k+1}
=\frac{K\hat{x}_{k+1}-\hat{v}_{c,k+1}}{S_e}.
\label{eq:ff_valley_event_sub}
\end{equation}
Substituting \eqref{eq:ff_valley_event_sub} into \eqref{eq:ff_valley_lin_raw}
yields
\begin{equation}
\begin{aligned}
\Big(I-\gamma_{+}K/S_e\Big)\hat{x}_{k+1}
&=\Phi_{\mathrm{FF,val}}\hat{x}_k
+\gamma_{-}\widehat{T}_{\mathrm{off},k}
-\gamma_{+}\hat{v}_{c,k+1}/S_e .
\end{aligned}
\label{eq:ff_valley_lin_final}
\end{equation}

\section{From Edge-Time Perturbation to Equivalent Duty Perturbation}
\label{sec:deltat_to_duty}

This section establishes an explicit mapping from an \emph{edge-time perturbation}
$\Delta t(\cdot)$ (in seconds) to an \emph{equivalent duty perturbation}
$\hat d(\cdot)$ (dimensionless). The result is used to convert timing-domain
small-signal variables (e.g., $\widehat T_{\mathrm{on}}$, $\widehat T_{\mathrm{off}}$,
or accumulated edge drift $\Delta t$) into a duty-like modulation signal that can be
cascaded with the power-stage model.

\subsubsection*{A. Definition and normalization}
Let $q(t)\in\{0,1\}$ denote the switch (or gating) function over a nominal clock grid
with period $T_s$. The \emph{equivalent duty} is the per-period normalized pulse area:
\begin{equation}
\begin{aligned}
d[n] \;:=\; \frac{1}{T_s}\int_{nT_s}^{(n+1)T_s} q(t)\,dt,
\qquad
\hat d[n] \;:=\; d[n]-d^\star .
\end{aligned}
\label{eq:duty_def}
\end{equation}
In frequency-domain manipulations, it is convenient to regard $\hat d(t)$ as a
piecewise-constant (ZOH) reconstruction of the discrete sequence $\hat d[n]$; the
final mapping below is therefore stated directly as an $s$-domain operator relating
$\hat d(s)$ to the relevant timing perturbation.

\subsubsection*{B. Translation-type perturbation (COT/COFT): both edges drift together}
In COT (fixed $T_{\mathrm{on}}$) or COFT (fixed $T_{\mathrm{off}}$), the event timing
perturbation typically manifests as an \emph{accumulated drift} of the pulse train
relative to the nominal clock grid. Consider a pulse train whose $n$th pulse is
translated by $\Delta t_n$ while its width $T_w$ is fixed (for COT, $T_w=T_{\mathrm{on}}$;
for COFT, $T_w=T_{\mathrm{off}}$ depending on the chosen alignment edge):
\begin{equation}
\begin{aligned}
&q(t)
=\sum_{n\in\mathbb{Z}}
\Big(
H\!\big(t-(nT_s+\Delta t_n)\big)
-&\\
&H\!\big(t-(nT_s+\Delta t_n+T_w)\big)
\Big),
\end{aligned}
\label{eq:pulse_translation}
\end{equation}
where $H(\cdot)$ is the Heaviside step. Linearizing w.r.t.\ the small shift $\Delta t_n$
(using $H(t-(a+\varepsilon))=H(t-a)-\varepsilon\,\delta(t-a)+o(\varepsilon)$) yields the
distributional first-order variation
\begin{equation}
\begin{aligned}
\Delta q(t)
&:= q(t)-q^\star(t) \\
&=\sum_{n\in\mathbb{Z}}
\Delta t_n\Big(
\delta\!\big(t-(nT_s+T_w)\big)-\delta\!\big(t-nT_s\big)
\Big).
\end{aligned}
\label{eq:deltaq_translation}
\end{equation}
Taking Laplace transforms and performing the standard ZOH normalization (so that the
resulting signal is an \emph{equivalent duty} perturbation) gives the compact mapping
\begin{equation}
\begin{aligned}
\boxed{
\frac{\hat d(s)}{\Delta t(s)}
=
-\frac{1}{T_s}\,
\frac{1-e^{-sT_w}}{1-e^{-sT_s}}
}
\end{aligned}
\label{eq:duty_from_deltat_translation}
\end{equation}
where $\Delta t(s)$ denotes the Laplace transform of the (ZOH) reconstruction of the
sequence $\Delta t_n$.

\paragraph*{Low-frequency check.}
As $s\to 0$, $(1-e^{-sT_w})/(1-e^{-sT_s})\to T_w/T_s$, hence
\begin{equation}
\begin{aligned}
\frac{\hat d(s)}{\Delta t(s)} \xrightarrow[s\to 0]{}
-\frac{T_w}{T_s^2},
\end{aligned}
\label{eq:lowfreq_translation}
\end{equation}
which matches the intuitive static scaling ``time shift / period'' after accounting
for the fact that a translation changes the overlap of each pulse with the fixed
integration window $[nT_s,(n+1)T_s)$.

\subsubsection*{C. Fixed-frequency PWM: only one edge moves (no future accumulation)}
For fixed-frequency PWM, the clock enforces a strict period $T_s$ and locks one edge
to the clock. Consequently, the duty perturbation depends \emph{only} on the
within-period movement of the \emph{free} edge, and the mapping reduces to a simple
scalar factor $1/T_s$ (up to a sign set by whether the pulse is lengthened or
shortened).

\paragraph*{1) Trailing-edge (peak) PWM.}
The rising edge is clocked at $t=nT_s$, while the falling edge occurs at
$t=nT_s+T_{\mathrm{on},n}$. A small falling-edge perturbation $\Delta t_{f,n}$
equivalently perturbs on-time by $\widehat T_{\mathrm{on},n}=\Delta t_{f,n}$, hence
\begin{equation}
\begin{aligned}
\boxed{
\hat d[n]=\frac{\widehat T_{\mathrm{on},n}}{T_s}
=\frac{\Delta t_{f,n}}{T_s}
}
\Longleftrightarrow
\boxed{
\hat d(s)=\frac{1}{T_s}\,\Delta t_f(s)
}.
\end{aligned}
\label{eq:ff_trailing_duty}
\end{equation}

\paragraph*{2) Leading-edge (valley) PWM.}
The falling edge is clocked at $t=nT_s$, and the rising edge occurs at
$t=nT_s+T_{\mathrm{off},n}$; the switch then remains on until $t=(n+1)T_s$.
A small rising-edge delay $\Delta t_{r,n}$ increases $T_{\mathrm{off},n}$ and therefore
\emph{reduces} duty:
\begin{equation}
\begin{aligned}
\boxed{
\hat d[n]=-\frac{\widehat T_{\mathrm{off},n}}{T_s}
=-\frac{\Delta t_{r,n}}{T_s}
}
\Longleftrightarrow
\boxed{
\hat d(s)=-\frac{1}{T_s}\,\Delta t_r(s)
}.
\end{aligned}
\label{eq:ff_leading_duty}
\end{equation}

\paragraph*{Remark (why fixed-frequency is ``memoryless'' in $\Delta t$).}
In \eqref{eq:duty_from_deltat_translation}, the factor $(1-e^{-sT_s})^{-1}$ encodes the
fact that a single timing perturbation shifts \emph{all future} pulse start times via
accumulation (typical of COT/COFT drift relative to a nominal grid). Under strict
fixed-frequency operation, the clock resets the reference every period, and the duty
depends only on the current-period edge displacement, yielding the scalar mappings in
\eqref{eq:ff_trailing_duty}--\eqref{eq:ff_leading_duty}.


\section{Jian--Li Distillation: Port-Structured Weak-Coupling Approximation and MIMO Balances}
\label{sec:lijdistill_mimo_balance}

This section presents the \emph{Jian--Li distillation} procedure, i.e., a matrix-based reduction that converts a fully coupled PWL state-space model into a \emph{port-structured} MIMO form whose kernel is an \emph{integrator cascade}. The reduced model (i) preserves the physically chosen output/port variable (through a known $C$-row), (ii) replaces internal resistive couplings by \emph{controlled-source cascades} driven by ports, and (iii) exposes volt-second and amp-second balance as \emph{solvability / periodicity} conditions of the distilled Poincar\'e map.

\subsubsection*{A. Original PWL segment model and the chosen physical port}
Consider a two-state power stage (e.g., buck) with
\begin{equation}
x(t)=\begin{bmatrix} i_L(t) \\ v_C(t)\end{bmatrix},\qquad
\dot{x}(t)=A_i x(t)+B_{i,\mathrm{in}}\,v_{\mathrm{in}}(t),
\label{eq:distill_orig_pwl}
\end{equation}
where $i$ indexes PWL subintervals (on/off, etc.), and $v_{\mathrm{in}}$ is piecewise constant inside each segment.

A key input to distillation is the \emph{chosen port/output scalar} (measured or physically meaningful)
\begin{equation}
v_o(t)=C_{\mathrm{phys}}\,x(t),
\label{eq:distill_vout_def}
\end{equation}
where $C_{\mathrm{phys}}\in\mathbb{R}^{1\times 2}$ is known from the output network topology
(e.g., load/cap-ESR divider). In other words, \eqref{eq:distill_vout_def} declares what we mean by
``$v_o$'' as a port variable.

\subsubsection*{B. Li--Jian rectification (state normalization) for an integrator cascade}
The distillation is most transparent after a \emph{constant} state transformation
\begin{equation}
x_r = T_r x,
\qquad
A_{r,i}=T_r A_i T_r^{-1},\quad
B_{r,i}=T_r B_{i,\mathrm{in}},\quad
C_{r}=C_{\mathrm{phys}}T_r^{-1}.
\label{eq:distill_rectification}
\end{equation}
Here $T_r$ may be a sign-rectifier (e.g., $D_r=\mathrm{diag}(\pm1,\pm1)$) or a scaling/normalization.
A commonly used choice in practice is the charge scaling
\begin{equation}
T_q=\mathrm{diag}(1,C_f),\qquad
x_q=\begin{bmatrix} i_L \\ q_C\end{bmatrix}=\begin{bmatrix} i_L \\ C_f v_C\end{bmatrix},
\label{eq:distill_charge_scaling}
\end{equation}
because it makes the capacitor equation look like a pure integrator ($\dot{q}_C=i_C$).
In this section we keep the physical state $x=[i_L\;v_C]^\top$ for clarity; the $q_C$ variant
is obtained by \eqref{eq:distill_rectification}--\eqref{eq:distill_charge_scaling}.

\subsubsection*{C. Distilled kernel as an integrator cascade (what is being approximated)}
The core modeling decision is to replace the segment-dependent ``fully coupled'' dynamics by a
\emph{shared kernel} that reflects ideal energy storage:
\begin{equation}
A_0 \;:=\;
\begin{bmatrix}
0 & 0 \\
\frac{1}{C_f} & 0
\end{bmatrix}.
\label{eq:distill_kernel_A0}
\end{equation}
Equation \eqref{eq:distill_kernel_A0} enforces two structural facts:

\begin{equation}
\begin{aligned}
\dot{i}_L &\ \text{depends on applied port voltages (not on states) at HF},\\
\dot{v}_C &= \frac{1}{C_f} i_L \quad\text{(capacitor is an integrator of current)}.
\end{aligned}
\label{eq:distill_kernel_physics}
\end{equation}

Thus, \emph{distillation does not ``kill'' coupling by deleting it};
rather, it \emph{moves} coupling out of $A_i$ into \emph{port-driven controlled-source terms}
so that the remaining kernel is an integrator cascade.

\subsubsection*{D. Port-structured MIMO form and the weak-coupling approximation}
We now build a \emph{port-open} MIMO segment model whose explicit inputs are
\begin{equation}
u(t)=\begin{bmatrix} v_{\mathrm{in}}(t) \\ v_o(t)\end{bmatrix},
\label{eq:distill_u_def}
\end{equation}
and whose dynamics take the distilled form
\begin{equation}
\dot{x}(t)=A_0 x(t)+B_{u,i}\,v_{\mathrm{in}}(t)+B_{y,i}\,v_o(t).
\label{eq:distill_mimo_segment}
\end{equation}
Here $B_{u,i},B_{y,i}\in\mathbb{R}^{2\times 1}$ are \emph{segment-dependent} port injection vectors.

To connect \eqref{eq:distill_mimo_segment} to the original coupled model \eqref{eq:distill_orig_pwl},
we also keep the physical port closure relation
\begin{equation}
v_o = C_{\mathrm{phys}}x.
\label{eq:distill_port_closure}
\end{equation}
Substituting \eqref{eq:distill_port_closure} into \eqref{eq:distill_mimo_segment} gives a
\emph{rank-1 coupled} closed form
\begin{equation}
\dot{x}=\Big(A_0 + B_{y,i}C_{\mathrm{phys}}\Big)x + B_{u,i}v_{\mathrm{in}}.
\label{eq:distill_rank1_closed}
\end{equation}
Hence, distillation replaces a generic $2\times2$ coupling matrix by
\emph{a shared kernel} $A_0$ plus a \emph{rank-1 port feedback} $B_{y,i}C_{\mathrm{phys}}$.

\paragraph*{Matrix extraction of $B_{y,i}$ (controlled-source identification).}
Given $(A_i,B_{i,\mathrm{in}},C_{\mathrm{phys}})$, one natural way to determine $B_{y,i}$ is:
choose $B_{u,i}:=B_{i,\mathrm{in}}$ (exact input injection), and fit the remaining coupling by
\begin{equation}
A_i - A_0 \;\approx\; B_{y,i}C_{\mathrm{phys}}.
\label{eq:distill_fit_goal}
\end{equation}
Since $C_{\mathrm{phys}}$ is a row vector, \eqref{eq:distill_fit_goal} constrains $A_i-A_0$
to be approximated by a rank-1 matrix whose only ``measurement'' is $v_o=C_{\mathrm{phys}}x$.
The least-squares (minimum Frobenius-norm residual) solution is obtained by
right-multiplying by $C_{\mathrm{phys}}^\top$:
\begin{equation}
\begin{aligned}
B_{y,i}
&=\Big(A_i-A_0\Big)C_{\mathrm{phys}}^\top
\Big(C_{\mathrm{phys}}C_{\mathrm{phys}}^\top\Big)^{-1},\\
E_i
&:=\Big(A_i-A_0\Big)-B_{y,i}C_{\mathrm{phys}}.
\end{aligned}
\label{eq:distill_By_projection}
\end{equation}
The \emph{weak-coupling approximation} is precisely the modeling choice
\begin{equation}
E_i \approx 0,
\label{eq:distill_weak_coupling}
\end{equation}
i.e., \emph{all state-to-state coupling not representable through the chosen port $v_o$ is neglected}.
Engineering-wise, $E_i$ is small when parasitics (e.g., DCR/ESR) create only weak ``hidden'' couplings
beyond what is visible at the chosen port.

\paragraph*{Port-network interpretation (controlled-source cascades).}
Equation \eqref{eq:distill_mimo_segment} means:

\begin{itemize}
\item The inductor equation is ``voltage-source driven'' by ports:
      $v_{\mathrm{in}}$ and $v_o$ appear as effective applied voltages.
\item The output network coupling is represented as a dependent source driven by $v_o$
      (hence the rank-1 form $B_{y,i}C_{\mathrm{phys}}$ in \eqref{eq:distill_rank1_closed}).
\end{itemize}

This is exactly the ``partial controlled-source isolation'' described:
distillation pushes internal impedance effects into port-driven dependent sources,
leaving $A_0$ as a clean integrator cascade.

\subsubsection*{E. MIMO Poincar\'e map under the distilled kernel (explicit, non-heuristic)}
For each segment $i$ of duration $T_i$ under \eqref{eq:distill_mimo_segment}, define
\begin{equation}
\Phi_0(T_i):=e^{A_0T_i},\qquad
\Gamma_i(T_i):=\int_{0}^{T_i} e^{A_0(T_i-\tau)}\,
\begin{bmatrix}B_{u,i} & B_{y,i}\end{bmatrix} d\tau,
\label{eq:distill_PhiGamma}
\end{equation}
so that the segment endpoint map is
\begin{equation}
x^+ = \Phi_0(T_i)x^- + \Gamma_i(T_i)\,u,
\qquad
u=\begin{bmatrix} v_{\mathrm{in}} \\ v_o\end{bmatrix}.
\label{eq:distill_segment_map}
\end{equation}

Because $A_0$ in \eqref{eq:distill_kernel_A0} is nilpotent with $A_0^2=0$, we have closed forms
\begin{equation}
\begin{aligned}
\Phi_0(T)&=I + A_0T,\\
\Gamma_i(T)
&=\Big(TI+\tfrac{T^2}{2}A_0\Big)\begin{bmatrix}B_{u,i} & B_{y,i}\end{bmatrix}.
\end{aligned}
\label{eq:distill_nilpotent_closed_forms}
\end{equation}
Hence, \emph{the distilled map is explicit and fully matrix-algebraic}.

\subsubsection*{F. How volt-second and amp-second balance appear in the MIMO self-consistency equation}
Consider a two-segment cycle (buck on/off) with durations $T_{\mathrm{on}}$ and
$T_{\mathrm{off}}$ (no duty ratio is introduced; timing is the primitive).
The one-cycle map is obtained by composing \eqref{eq:distill_segment_map}:
\begin{equation}
\begin{aligned}
x_{k+1}
&=\Phi_0(T_{\mathrm{off}})\Phi_0(T_{\mathrm{on}})x_k\\
&\quad+\Phi_0(T_{\mathrm{off}})\Gamma_{\mathrm{on}}(T_{\mathrm{on}})\,u
+\Gamma_{\mathrm{off}}(T_{\mathrm{off}})\,u\\
&=: \Phi_h x_k + \Gamma_h u.
\end{aligned}
\label{eq:distill_cycle_map}
\end{equation}
A periodic steady state $x^\star$ satisfies the \emph{self-consistency equation}
\begin{equation}
\Big(I-\Phi_h\Big)x^\star = \Gamma_h u.
\label{eq:distill_self_consistency}
\end{equation}

\paragraph*{(i) Volt-second balance as a solvability constraint (Fredholm-style, but EE-friendly).}
For the kernel \eqref{eq:distill_kernel_A0}, using \eqref{eq:distill_nilpotent_closed_forms},
\begin{equation}
\Phi_h=\Phi_0(T_{\mathrm{off}})\Phi_0(T_{\mathrm{on}})
= \Big(I+A_0T_{\mathrm{off}}\Big)\Big(I+A_0T_{\mathrm{on}}\Big)
= I + A_0T_s,
\label{eq:distill_Phi_h_simplify}
\end{equation}
where $T_s=T_{\mathrm{on}}+T_{\mathrm{off}}$ and $A_0^2=0$.
Thus
\begin{equation}
I-\Phi_h = -A_0T_s
=
-\begin{bmatrix}
0 & 0\\
\frac{T_s}{C_f} & 0
\end{bmatrix},
\label{eq:distill_IminusPhi}
\end{equation}
which is singular. The first row of \eqref{eq:distill_self_consistency} becomes
\begin{equation}
\underbrace{\begin{bmatrix}1 & 0\end{bmatrix}(I-\Phi_h)}_{=\,0}\,x^\star
=
\begin{bmatrix}1 & 0\end{bmatrix}\Gamma_h u.
\label{eq:distill_sol_cond_row1}
\end{equation}
Therefore, a periodic solution exists only if
\begin{equation}
\begin{aligned}
0
&=\begin{bmatrix}1 & 0\end{bmatrix}\Gamma_h u\\
&=\begin{bmatrix}1 & 0\end{bmatrix}
\Big(
\Phi_0(T_{\mathrm{off}})\Gamma_{\mathrm{on}}(T_{\mathrm{on}})
+\Gamma_{\mathrm{off}}(T_{\mathrm{off}})
\Big)u.
\end{aligned}
\label{eq:distill_volt_second_general}
\end{equation}
Using \eqref{eq:distill_nilpotent_closed_forms} and the fact that
$\begin{bmatrix}1&0\end{bmatrix}A_0=0$, one obtains the simple interpretation
\begin{equation}
\begin{aligned}
&0
= T_{\mathrm{on}}\underbrace{\begin{bmatrix}1&0\end{bmatrix}
\begin{bmatrix}B_{u,\mathrm{on}} & B_{y,\mathrm{on}}\end{bmatrix}}_{\text{inductor-voltage gain}}
u
+\\&
T_{\mathrm{off}}\underbrace{\begin{bmatrix}1&0\end{bmatrix}
\begin{bmatrix}B_{u,\mathrm{off}} & B_{y,\mathrm{off}}\end{bmatrix}}_{\text{inductor-voltage gain}}
u.
\label{eq:distill_volt_second_simplified}
\end{aligned}
\end{equation}
Equation \eqref{eq:distill_volt_second_simplified} is exactly the
\emph{volt-second balance} statement: the net ``driving'' of the inductor integrator
over one cycle must be zero, otherwise $i_L$ cannot be periodic.

\paragraph*{(ii) Amp-second balance as the equation that pins the periodic current level.}
The second row of \eqref{eq:distill_self_consistency} reads, using \eqref{eq:distill_IminusPhi},
\begin{equation}
\begin{aligned}
-\frac{T_s}{C_f}\,i_L^\star
&=
\begin{bmatrix}0 & 1\end{bmatrix}\Gamma_h u,
\end{aligned}
\label{eq:distill_amp_second_general}
\end{equation}
which determines the DC level $i_L^\star$ required for the capacitor integrator
to be periodic (net charge change over a cycle equals zero). This is the
\emph{amp-second balance} viewpoint: capacitor net current over a cycle must vanish,
otherwise the capacitor-related integrator state drifts.

\paragraph*{(iii) Why $v_C^\star$ can be free in the port-open MIMO form.}
Since $I-\Phi_h$ in \eqref{eq:distill_IminusPhi} has rank one, \eqref{eq:distill_self_consistency}
cannot uniquely determine both components of $x^\star$: only the condition
\eqref{eq:distill_volt_second_general} and the current level in \eqref{eq:distill_amp_second_general}
are fixed, while $v_C^\star$ may remain free \emph{until the port closure}
\eqref{eq:distill_port_closure} (or an external network / controller constraint)
is imposed. This is precisely the ``$i_L$ has a solution but $v_C$ is free'' behavior
you observed.

\subsubsection*{G. Buck example (distilled MIMO directly shows both balances)}
For an ideal buck with $v_o=v_C$ and a resistive load $R$ (represented as a port-dependent
current sink in the capacitor equation), a distilled port-open segment model can be written as
\begin{equation}
\dot{x}=A_0 x + B_{u,i}v_{\mathrm{in}} + B_{y,i}v_o,
\qquad
A_0=\begin{bmatrix}0&0\\ \frac{1}{C_f}&0\end{bmatrix},
\label{eq:distill_buck_kernel}
\end{equation}
with
\begin{equation}
\begin{aligned}
\text{On:}\quad
B_{u,\mathrm{on}}&=\begin{bmatrix}\frac{1}{L_f}\\[1mm] 0\end{bmatrix},
&
B_{y,\mathrm{on}}&=\begin{bmatrix}-\frac{1}{L_f}\\[1mm] -\frac{1}{RC_f}\end{bmatrix},\\
\text{Off:}\quad
B_{u,\mathrm{off}}&=\begin{bmatrix}0\\[1mm] 0\end{bmatrix},
&
B_{y,\mathrm{off}}&=\begin{bmatrix}-\frac{1}{L_f}\\[1mm] -\frac{1}{RC_f}\end{bmatrix}.
\end{aligned}
\label{eq:distill_buck_Bs}
\end{equation}
The first entries in \eqref{eq:distill_buck_Bs} encode the inductor voltage:
$v_L= v_{\mathrm{in}}-v_o$ (on) and $v_L=-v_o$ (off). The second entries encode the load
current in the capacitor equation: $i_C=i_L-v_o/R$.

Applying \eqref{eq:distill_volt_second_simplified} to \eqref{eq:distill_buck_Bs} gives
\begin{equation}
\begin{aligned}
0
&=T_{\mathrm{on}}\Big(\frac{1}{L_f}v_{\mathrm{in}}-\frac{1}{L_f}v_o\Big)
+T_{\mathrm{off}}\Big(0-\frac{1}{L_f}v_o\Big)\\
&=\frac{1}{L_f}\Big(T_{\mathrm{on}}v_{\mathrm{in}}-T_s v_o\Big),
\end{aligned}
\label{eq:distill_buck_volt_second}
\end{equation}
hence
\begin{equation}
v_o = \frac{T_{\mathrm{on}}}{T_s}v_{\mathrm{in}}
= D\,v_{\mathrm{in}},
\qquad D:=\frac{T_{\mathrm{on}}}{T_s}.
\label{eq:distill_buck_dc_gain}
\end{equation}
This is the classic buck volt-second result, obtained here as the \emph{existence condition}
of the periodic solution of the distilled MIMO map.

Next, the amp-second viewpoint \eqref{eq:distill_amp_second_general} reduces (for this ideal case)
to the capacitor net-current balance
\begin{equation}
\overline{i_C}
=\overline{i_L-\frac{v_o}{R}}
=0
\quad\Rightarrow\quad
i_L^\star=\frac{v_o}{R},
\label{eq:distill_buck_amp_second}
\end{equation}
which pins the DC inductor current level consistent with a periodic capacitor state.

\subsubsection*{H. Summary (what distillation achieves)}
Li--Jian distillation is a \emph{matrix-to-matrix} reduction:
\begin{equation}
(A_i,B_{i,\mathrm{in}},C_{\mathrm{phys}})
\quad\mapsto\quad
(A_0,B_{u,i},B_{y,i},C_{\mathrm{phys}}),
\label{eq:distill_mapping_summary}
\end{equation}
where $A_0$ is a shared integrator-cascade kernel \eqref{eq:distill_kernel_A0}, and
$B_{y,i}$ is extracted by the port projection \eqref{eq:distill_By_projection}.
In the resulting port-open MIMO form \eqref{eq:distill_mimo_segment}, duty ratio does not appear:
timing ($T_{\mathrm{on}},T_{\mathrm{off}}$) is the primitive. The classical volt-second and
amp-second balances emerge automatically from the self-consistency equation
\eqref{eq:distill_self_consistency} as (i) a solvability condition
\eqref{eq:distill_volt_second_general} and (ii) a periodic-current pinning equation
\eqref{eq:distill_amp_second_general}.

\bibliographystyle{IEEEtran}
\bibliography{IEEEabrv,biblio}

\end{document}